\documentclass[useAMS,usenatbib]{mn2e}
\usepackage{graphicx}

\title[Implementation of an Optimised Cassegrain System for Radio Telescopes]{Implementation of an Optimised Cassegrain System for Radio Telescopes}
\author[C. M. Holler et al]{C. M. Holler$^{1}$\thanks{Now at Sub-Dept of Astrophysics, University of Oxford,
			Denys Wilkinson Bldg, Keble Road, Oxford, OX1 7RH,
			United Kingdom}, R. E.
Hills$^{1}$, M. E. Jones$^{1}$\footnotemark[1], K. Grainge$^{1}$ and T. Kaneko$^{1}$\\
$^{1}$Cavendish Laboratory, Cambridge University, 
	      Cambridge CB3 0HE, United Kingdom}
\begin{document}

\date{Accepted 2007 November 29. Received 2007 November 05; in original form 2007 May 01}

\pagerange{\pageref{firstpage}--\pageref{lastpage}} \pubyear{2007}

\maketitle

\label{firstpage}

\begin{abstract}
We present the antenna design for a radio interferometer, the Arcminute Microkelvin Imager, together with its beam pattern measurement. Our aim was to develop a low-cost system with high aperture efficiency and low ground-spill across the frequency range 12--18~GHz. We use a modified cassegrain system consisting of a commercially-available paraboloidal primary mirror with a diameter of 3.7~m, and a shaped secondary mirror. The secondary mirror is oversized with respect to a ray-optics design and has a surface that is bent towards the primary near its outer edge using a square term for the shaping. The antennas are simple to manufacture and therefore their cost is low. The design increased the antenna gain by approximately 10~per~cent compared to a normal Cassegrain system while still maintaining low contamination from ground-spill and using a simple design for the horn.
\end{abstract}


\section{Introduction}

In radio astronomy, advances in front-end receiver technology and increased bandwidths have steadily improved the sensitivity of radio telescopes. Nevertheless, observations of weak radio signals are difficult and the signals are easily contaminated by strong radio sources outside the main beam and emissions from the ground. For close-packed radio interferometers, signals from other antennas in the array are an additional source of contamination. Although interferometers can provide excellent control over systematic errors, the development of a large array is costly due to the large number of individual antennas. Therefore it is essential to be able to produce antennas with high aperture efficiency and low ground-spill over a broad bandwidth at low cost.

A classical Cassegrain system (paraboloidal primary, hyperboloidal secondary) typically provides an aperture efficiency of around 70~per~cent with significant ground-spill. The well-known alternative, a shaped Cassegrain system, where both mirrors deviate from their classical shapes \citep{b1}, can give much higher efficiency and reduced ground-spill. But it is relatively costly, since both the primary and secondary mirrors must be specially shaped for the particular design. We therefore decided to develop an antenna using a commercially-available paraboloidal reflector as the primary element and to modify only the shape and size of the secondary mirror, as well as optimizing the position of the feed.  As we shall demonstrate, this works well for a system where the diameter of the secondary mirror, expressed in wavelengths, is not very large. 

The design implements and extends the ideas of  \cite{b8}, who showed that aperture efficiency of such systems can be increased by oversizing the secondary, and \citet{b7}, who used a flange to redirect rays from the oversized part to reduce spillover. The additional steps here are to optimise both the ground-spill as well as the efficiency and also to find a solution giving good performance over a wide bandwidth. An optimisation using physical optics/physical theory of diffraction PO/PTD analysis results in a square term for the shaping of the outer parts of the secondary, which starts well within its ray optics diameter. This shaping increases aperture efficiency further. The idea of a "partially shaped" Cassegrain has also been described by \citet{b2}, who showed that aperture efficiency can be increased by shaping just the secondary.

The antenna system presented here has been developed for the Arcminute Microkelvin Imager \citep{b4, b3} a ten element interferometric array designed to observe the imprints of distant clusters of galaxies in the cosmic microwave background radiation (CMB), the so-called Sunyaev--Zel'dovich (SZ) effect \citep{b5}. A similar optical design has also been adapted for the upgrade of the Cosmic Background Imager  \citep{b9, b10}, a 13 element interferometric array measuring the CMB power spectrum.

\section[]{Design Procedure}

\subsection{Oversized Secondary}\label{sec:ov_secondary}

\begin{figure}
 \centering
  \includegraphics[width=3.0in]{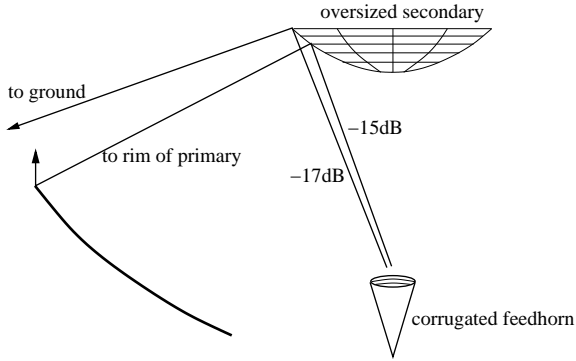}
  \caption{The principle of oversizing the secondary mirror.
\label{fig:corr_transm_refl}}
\end{figure}

The design starts with a normal Cassegrain system, which is modified in two steps. The first step is to oversize the secondary reflector, with respect to the size that would be chosen in a ray optics design, as shown in Fig.~\ref{fig:corr_transm_refl}.  (In this example the feed taper at the edge of the secondary would have been 15dB for a geometrically sized secondary, while it is 17dB for the oversized one.)

\begin{figure}
  \centering
  \includegraphics[width=3.3in]{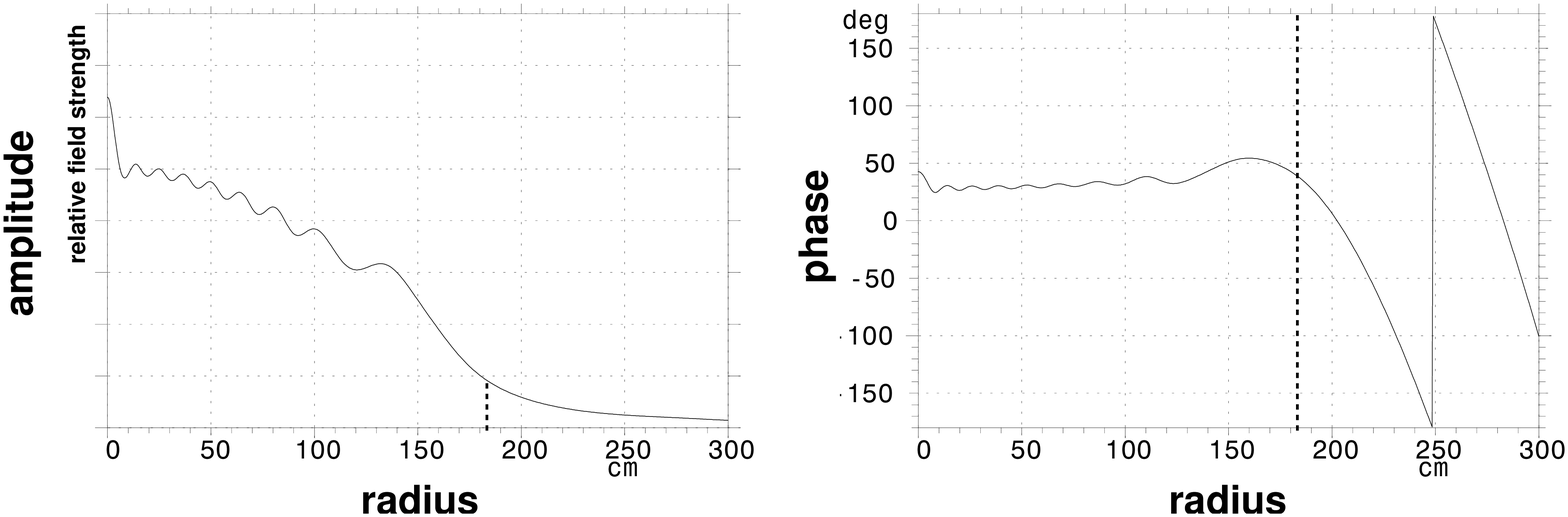}
  \includegraphics[width=3.3in]{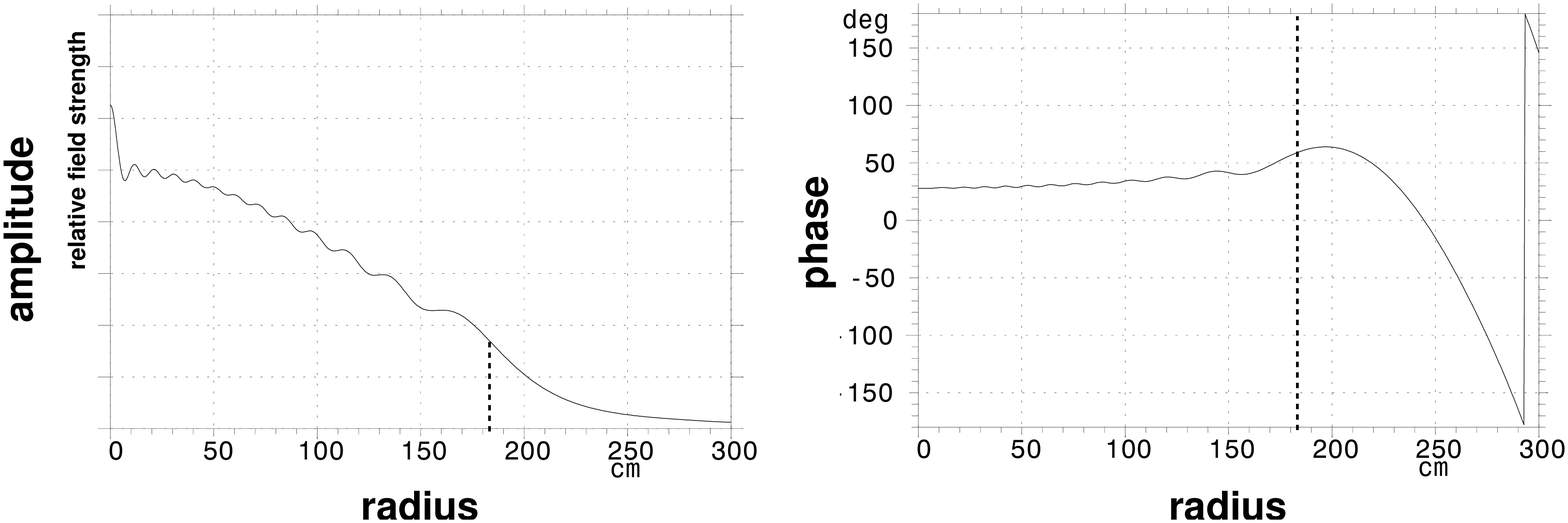}
  \caption{The aperture illumination of a normal Cassegrain system (top) and of a Cassegrain system with oversized secondary mirror (bottom). The vertical dotted lines at radius~185~cm mark the edge of the primary.
\label{fig:ap_ill_overs}}
\end{figure}

A physical optics calculation of this system shows that oversizing the secondary without changing the feed illumination increases the aperture efficiency. To understand this, we examine the field distributions on the primary mirror when illuminated by the secondary. Fig.~\ref{fig:ap_ill_overs} compares these for a normal Cassegrain and a Cassegrain with oversized secondary.

When the secondary mirror has the size determined by geometrical optics there is a local maximum in the amplitude of the illumination which is considerably inside the edge of the primary mirror. Oversizing the secondary moves this maximum further out and this produces an illumination which is more uniform across the reflector. For example, at 15~GHz and with a 3.7-m primary, we found that making the secondary mirror a factor of 1.2 larger than indicated by the geometry (to a diameter of 498~mm) increases the aperture efficiency from 0.73 to 0.77.

\begin{figure}
  \centering
  \includegraphics[width=3.2in]{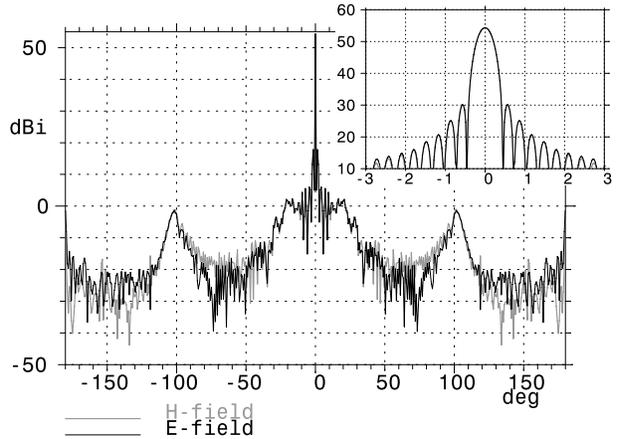}
  \caption{The simulated beam pattern of an oversized Cassegrain system at 15~GHz. The inset shows the main beam with an expanded scale.
\label{fig:overs_cass}}
\end{figure}

The disadvantage is the higher sidelobes due to spill-over at about 100$\degr$ from the main beam, directly pointing at the hot ground.  Fig.~\ref{fig:overs_cass} shows the full simulated beam pattern for such a system at the centre of our band (15~GHz).  The high far-out sidelobes are clearly visible. They are 55dB below  the peak and contain about 4.2~per~cent of the total power.

\subsection{Shaped-edge Secondary}\label{sec:ro-secondary}

\begin{figure}
  \centering
  \includegraphics[width=3.0in]{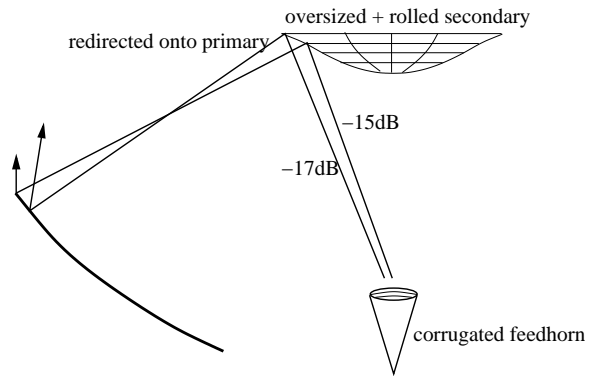}
  \caption{The concept of a shaped-edge Cassegrain system. Rays from the edge of the secondary are redirected onto the  primary mirror.
\label{fig:rolled_sec}}
\end{figure}

The second step of the design procedure is therefore aimed at reducing these far-out sidelobes, but also at increasing the aperture efficiency further. This is done by bending the outer section of the secondary reflector inwards (towards the primary). In geometric optics this will redirect rays that would otherwise point at the ground onto the primary, as shown in Fig.~\ref{fig:rolled_sec}. 

\begin{figure}
  \centering
  \includegraphics[width=3.3in]{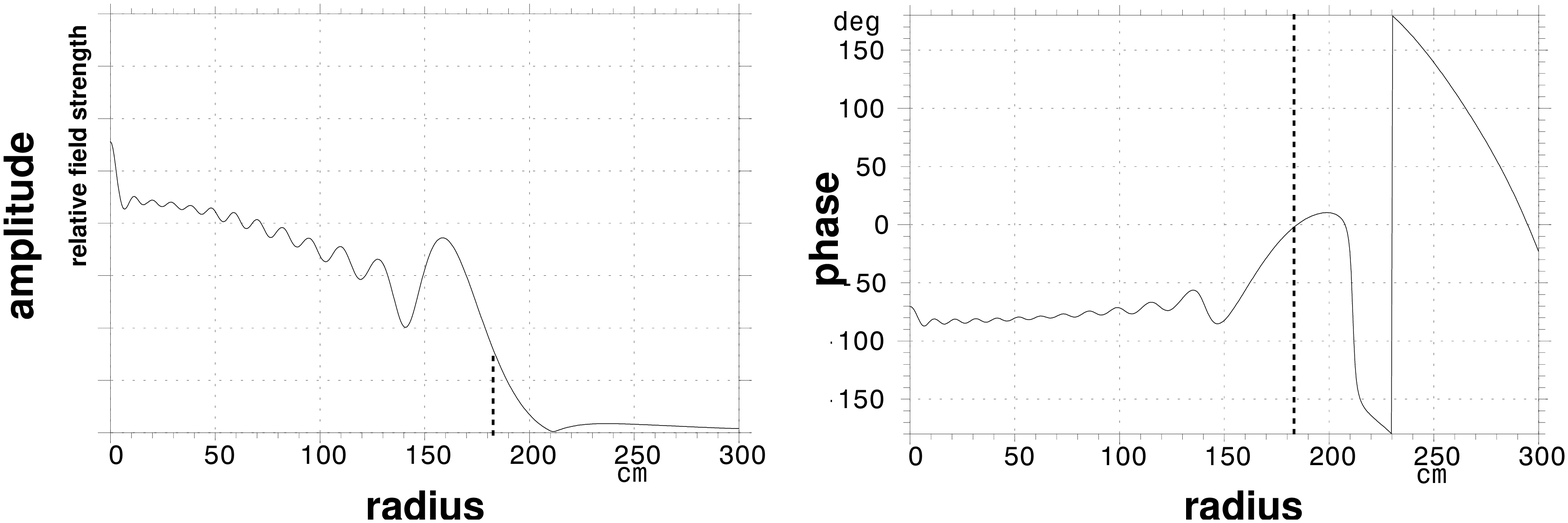}
  \caption{The aperture illumination of a shaped-edge Cassegrain system.
\label{fig:ap_ill_rolled}}
\end{figure}

It is clear from this diagram that these rays will then end up on the sky pointing in a direction that is close to that of the main beam, but not exactly the same.  In a diffraction analysis this difference in direction shows up as a phase change in the illumination pattern in the outer parts of the primary. At the relatively long wavelengths we are using here, however, this effect is not large, as is shown in Fig.~\ref{fig:ap_ill_rolled}. 

Comparison with Fig.~\ref{fig:ap_ill_overs} reveals that, although there are now rather larger ripples in the illumination, the overall effect is to increase the amplitude on the reflector with the result that the aperture efficiency is further increased to 0.80. 

Even with the geometry of the primary mirror fixed, there are now quite a large number of parameters required to describe the system. These include: the diameter of the secondary; the point at which the shaping begins;  the amount of deviation from the nominal hyperboloid which increases quadratically from that point;  and the diameter, length and position of the feed horn.  Although it would be possible in principle to do a fully automated search of this parameter space in order to find an optimum for all of these, there are sufficient practical constraints, e.g. on the size and length of the horn (which needs to fit inside a cryostat) that this was not thought to be necessary. A program based on scalar diffraction theory was however written which could analyse this system relatively quickly and produce estimates of the aperture efficiency and spill-over loss at three frequencies, 12.5, 15 and 17.5~GHz.  

This program consisted of:  First generating the near-field pattern of the corrugated feed on a spherical surface at the approximate position of the secondary;  then calculating the contributions from each small area of the secondary to the field at a series of points on a radial section across the primary (i.e. generating quantities of the type shown in Figures~\ref{fig:ap_ill_overs} and \ref{fig:ap_ill_rolled});  and finally, integrating these fields (assuming circular symmetry) to find the aperture efficiency and spill-over loss of the system. Although there are a number of approximations in this approach, it proved to be quite accurate as a first estimate of the optimum parameter space (mainly because the reflectors are many wavelengths across). In particular the changes in the efficiency for small changes in the parameters were found to be reliable.

The program included a routine to find the axial position of the feed to give the minimum phase error (weighted with the aperture illumination and averaged over the three frequencies).  This was done automatically for each variant because the gain is quite sensitive to this parameter.  The various other parameters listed above were then stepped through a range of values so that their effects could be understood.  This procedure was found to converge on a good solution without any great difficulty. The aperture efficiency and spill-over were well-behaved functions of the variables and so a design with good values of both could be found.

The chosen design was then subjected to a more rigorous analysis using the GRASP software package \citep{b6}. This provides a full electro-magnetic simulation using methods that are known to be accurate under the present circumstances. The analysis proceeds from the feed, to the secondary, to the primary, and then to the far field. The physical optics approximation is used to calculate the currents induced in each element by the fields radiated by the previous one. The next set of fields are then calculated by numerical integration of the radiation by these currents. The effects of the edges of the reflectors are treated by means of the physical theory of diffraction.

The final parameters for our design are a 3.7-m diameter primary mirror with a 498-mm diameter secondary mirror, which is oversized by a factor of 1.2 compared to nominal Cassegrain geometry; a shaping of the secondary mirror, which starts at $r=170\,\rm mm$  using a square term, with a maximum deviation from the hyperboloidal shape of 13~mm at the secondary edge; and a feed consisting of a corrugated horn with 6.16$\degr$ semi-flare angle and 108-mm aperture diameter.

\begin{figure}
  \centering
  \includegraphics[width=3.2in]{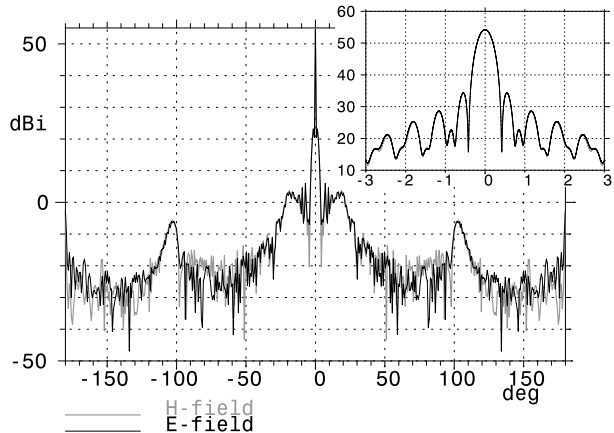}
  \caption{The simulated beam pattern of a shaped-edge Cassegrain system at 15~GHz. The insets show the main beam with an expanded scale. 
\label{fig:pattern12_rolled}}
\end{figure}

The beam pattern, showing close-in and far-out sidelobes, is presented in Fig.~\ref{fig:pattern12_rolled}. The insets of this Figure and of Fig.~\ref{fig:overs_cass} show the beam shapes in the forward direction and have been corrected for the shadowing due to the secondary mirror using a hole in the primary with equal diameter to the secondary in the simulations. 

The final results are summarised in Table~1. The results show an average of about 2~per~cent of power in the far-out sidelobes. This means that the contribution to the system temperature would be about 6K, if these sidelobes fell entirely on the ground at 300K. This is still an unacceptably large fraction of the expected system temperture for our purposes. To suppress sidelobes further we added a 15-cm wide metal shield mounted around the edge of the primary at an angle of 45$\degr$. This also has the additional benefit of reducing possible crosstalk between adjacent antennas of the array. This measure brings the amount of power within the far-out sidelobes (90$\degr$ to 120$\degr$) down to only 0.9, 0.3 and 0.2~per~cent at 12.5, 15 and 17.5~GHz, respectively.  

\begin{table}
\centering
\begin{tabular}{c|c c c}\hline
& $12.5\,\rm GHz$ & $15\,\rm GHz$ & $17.5\,\rm GHz$ \\ \hline 
\small{Aperture eff.} $\eta$ & $0.77$ & $0.80$ & $0.75$ \\ 
\small{Sidelobe level at $100^\circ$} & $-55.5\,\rm dB$ & $-60.5\,\rm
dB$ & $-66.5\,\rm dB$ \\ 
\small{Sidelobe power at $100^\circ$} & $3.4\%$ & $1.6\%$ & $0.6\%$  \\ 
\small{Beam eff. (within $\pm 1^\circ$)} & $0.74$&$0.84$&$0.82$ \\ 
\hline
$\eta$ \small{(normal Cass.)} & $0.72$ & $0.73$ & $0.71$ \\ 
$\eta$ \small{(oversized Cass.)}& $0.74$ & $0.77$ & $0.74$ \\ \hline
\end{tabular}\\ [0.5ex]
\caption{
Simulation results for the shaped-edge Cassegrain system using a 498-mm 
diameter shaped secondary reflector and a 3.7-m diameter
paraboloidal primary. As a comparison the
aperture efficiencies for the normal Cassegrain and the oversized secondary
Cassegrain are also given.}
\end{table}

\section{Beam Pattern Measurement}

The beam pattern of an antenna within an interferometer can be derived observationally from measuring the complex antenna voltage pattern using a single baseline. This is achieved by pointing one antenna permanently at a bright source and then scanning the other antenna across the source. The antenna power beam pattern is then given by the voltage pattern multiplied with its complex conjugate. Since the amplitude of the antenna voltage pattern falls only as the square root of the power beam pattern, it is possible to make precise measurements of the main beam and the first few sidelobes without being greatly limited by the noise level of the system. However, the far-out sidelobes at 100$\degr$ from the main beam are too low-level to be detected directly on this system.

\begin{figure}
  \centering
  \includegraphics[width=3.2in]{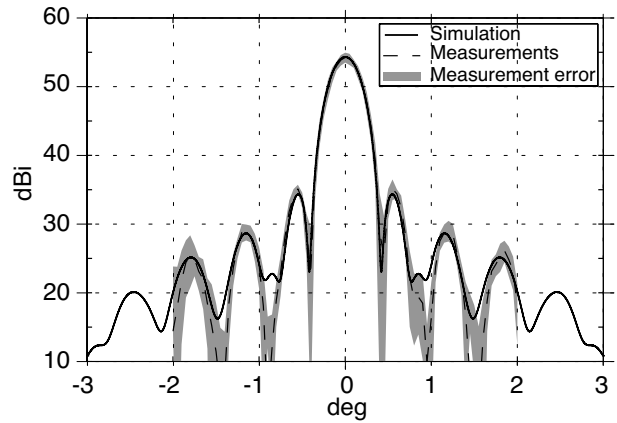}
  \caption{The simulated beam pattern at 15~GHz (solid line) compared with a measurement of the beam pattern at 15.4~GHz using the Arcminute Microkelvin Imager (dashed line). The grey area shows the variations of several measurements. 
\label{fig:meas_pattern}}
\end{figure}

The result of the beam pattern measurement at 15.4~GHz is shown in Fig.~\ref{fig:meas_pattern} and compared with the simulated pattern at 15.0~GHz. Note that for the figure the 15~cm metal shield around the primary mirror mentioned in Section~\ref{sec:ro-secondary} has been included in the simulation, which therefore differs slightly from the simulation in Fig.~\ref{fig:pattern12_rolled}. For our measurement we used  3C84, one of the brightest radio sources in the sky that is unresolved on the baseline being used. The curve was sampled at 0.05$\degr$ steps from -2$\degr$ to +2$\degr$ with a sample time of 4~minutes at each step. It shows excellent agreement with the simulation. The amplitudes of the curves have been aligned in the forward direction, since a measurement of an absolute antenna gain is very difficult.

\section{Conclusions}

We have presented the design, development and results of a Cassegrain optical system with a specially shaped secondary mirror, designed for the Arcminute Microkelvin Imager, using a low-cost 3.7~m commercial paraboloidal main reflector. This design gives high aperture efficiency with low ground-spill over a broad bandwidth. We have shown the changes necessary to modify the secondary mirror and the effect this has on the aperture illumination function on the primary reflector. In addition we have presented simulated antenna beam patterns for the centre of the frequency band at 15~GHz. These also show the increase in efficiency by about 10~per~cent relative to a normal Cassegrain system while maintaining low ground-spill. Finally, the comparison of the simulated data with measurements of the beam pattern shows excellent agreement. 

\section*{Acknowledgments}

The authors would like to thank Angela Taylor for the initial measurement of the antenna pattern and the other members of the AMI collaboration for their contributions to the development and testing of the antenna system.

\bsp

\label{lastpage}

\end{document}